\documentclass[pre,aps,10pt,superscriptaddress,amsmath,amssymb,showpacs,twocolumn,floatfix]{revtex4-1}

\usepackage[unicode]{hyperref}
\usepackage[dvipsnames]{xcolor}

\usepackage{bm}
\usepackage{graphicx}
\usepackage{subfigure}
\usepackage{adjustbox}

\def\vp{\varphi}

 \DeclareMathOperator{\tr}{tr}

\renewcommand{\Re}{\mathop\mathrm{Re}\nolimits}

\begin{document}

\title{Cooper pair splitting in ballistic ferromagnetic SQUIDs}

\author{P.~L.\ Stroganov}
\affiliation{Moscow Institute of Physics and Technology, 141700 Dolgoprudny, Russia}

\author{Ya.~V.\ Fominov}
\affiliation{L.~D.\ Landau Institute for Theoretical Physics RAS, 142432 Chernogolovka, Russia}
\affiliation{National Research University Higher School of Economics, 101000 Moscow, Russia}

\date{15 November 2017}

\begin{abstract}
We consider ballistic SQUIDs with spin filtering inside half-metallic ferromagnetic arms.
A singlet Cooper pair cannot pass through an arm in this case, so the Josephson current is entirely due to the Cooper pair splitting, with two electrons going to different interferometer arms. In order to elucidate the mechanisms of Josephson transport due to split Cooper pairs, we assume the arms to be single-channel wires in the short-junction limit.
Different geometries of the system (determined by the length of the arms and the phases acquired by quasiparticles during splitting between the arms) lead to qualitatively different behavior of the SQUID characteristics (the Andreev levels, the current-phase relation, and the critical Josephson current) as a function of two control parameters, the external magnetic flux and misorientation of the two spin filters. The current-phase relation can change its amplitude and shape, in particular, turning to a $\pi$-junction form or acquiring additional zero crossings.
The critical current can become a nonmonotonic function of the misorientation of the spin filters and the magnetic flux (on half of period). Periodicity with respect to the magnetic flux is doubled, in comparison to conventional SQUIDs.
\end{abstract}

\maketitle

\tableofcontents

\section{Introduction}
In the field of superconductivity, SQUID physics is a direction of great fundamental and applied importance \cite{Barone}. The operation of the SQUID is based on interference between Cooper pairs passing through two different arms that comprise the interferometer loop. The external magnetic flux threading the loop is a control parameter that determines the Josephson supercurrent carried by the device; the critical current is periodic with a period of flux quantum $\Phi_0 = \pi\hbar c/e$. The presence of two arms allows, at least in principle, processes in which a Cooper pair is split and the two electrons pass through different arms (recombining again to form a Cooper pair afterwards). The transport processes involving pair splitting generally lead to the critical current that depends on the flux with doubled periodicity, $2\Phi_0$ \cite{Choi2000,Wang2011,Jacquet2015,Mironov2015}.

However, the split-pair processes become noticeable only when the passage of nonsplit Cooper pairs is suppressed. The role of splitting can be enforced by energy filtering or spin filtering inside the arms (by means of electrically tunable quantum dots or ferromagnetic filters, respectively). Correlated transport through the arms of such multiterminal devices has been studied both theoretically \cite{Recher2001,Lesovik2001,Borlin2002,Chtchelkatchev2002} and experimentally \cite{Hofstetter2009,Herrmann2010,Schindele2012,Das2012}. Josephson current through the SQUID with two arms containing quantum dots (where the Coulomb energy impedes passage of nonsplit Cooper pairs through each arm) was theoretically studied in Refs.~\cite{Choi2000,Wang2011,Jacquet2015,Probst2016} and recently realized experimentally \cite{Deacon2015}.

While the main attention up to now has been paid to the quantum-dot scheme of Cooper pair splitting, the ferromagnetic realization \cite{Lesovik2001,Chtchelkatchev2002} has certain advantages. Half-metallic (H) ferromagnets (already employed in various superconducting hybrid structures, see, e.g., Refs.\ \cite{Keizer2006,Anwar2010,Singh2015}) should lead to highly efficient splitting due to absolute spin filtering (we imply singlet superconductors with opposite spins of electrons in a Cooper pair and do not consider spin-active interfaces, which could lead to singlet-triplet conversion \cite{Eschrig2008}).
At the same time, mutual orientation of magnetizations in the two arms can be varied by a weak external magnetic field due to different (geometry-related) coercivities of the ferromagnetic filters or due to exchange bias applied to one of the arms (it has been demonstrated that the latter method can be used for continuous variation of the misorientation angle \cite{Leksin2012}).
This provides an additional degree of freedom for controlling the device.

The dependence of Cooper pair splitting on the angle between magnetizations of the arms has been studied experimentally in an SF setup \cite{BWL2004,BL2005}. Two ferromagnetic arms F were contacted to a superconductor S close to each other and a voltage was applied to one of the arms. This produced a current in the other ferromagnetic arm due to crossed Andreev reflection (which is just another side of Cooper pair splitting). The current was sensitive to the relative orientation of the magnetizations.

The Josephson effect in SQUIDs with magnetic arms has previously been theoretically considered in the diffusive limit \cite{Melin2005,Melin_comment,IOFF}. In this case, the disorder-averaged Josephson current is strongly suppressed due to different phases acquired by electrons passing through different arms (the difference is caused by different disorder configuration in the arms) \cite{Melin2003}. At the same time, in any particular sample this suppression is absent, so the current through the system is mainly due to mesoscopic fluctuations.

\begin{figure}[t]
\includegraphics[width=\columnwidth]{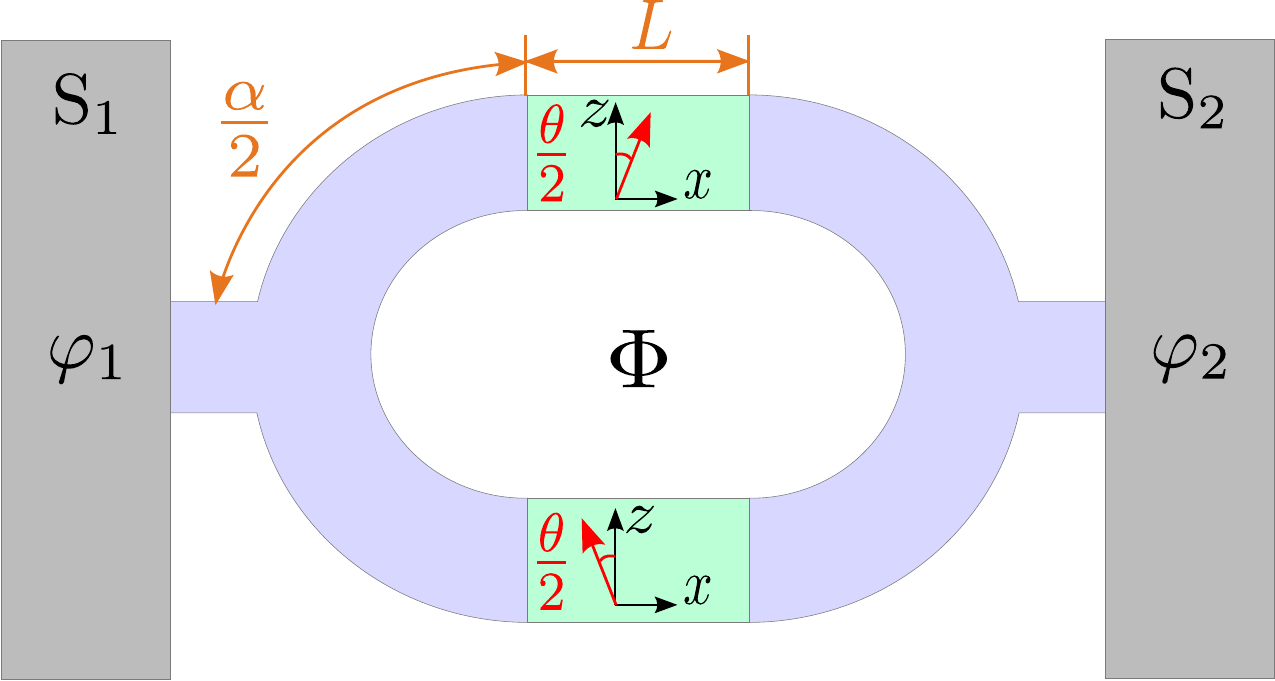}
\caption{Sketch of the ferromagnetic spin-filtering SQUID with splitting of Cooper pairs. Gray areas S$_{1,2}$ are singlet superconducting reservoirs, blue parts are three-terminal beam splitters (normal wires), and green insets are half-metallic ferromagnets (spin filters). Magnetizations of the filters (red arrows) lie in the $xz$ plane and are chosen to be symmetric with respect to $z$ axis, with angle $\theta$ between them. The interferometer loop is threaded by the external magnetic flux $\Phi$. The quantities $\alpha/2$ and $L$ are the phases accumulated by quasiparticles when passing the segments of the beam splitter and the (half-metallic) ferromagnetic insets.}
\label{fig: model_sketch}
\end{figure}

In this paper, we theoretically study the magnetic spin-filtering SQUID with splitting of the Cooper pairs in the opposite limit, when the arms are ballistic.
We consider the system schematically depicted in Fig.~\ref{fig: model_sketch}. Two beam splitters provide the possibility of Cooper pair splitting processes, and two half-metallic inserts (filters) block non-split processes (in which a Cooper pair could travel through a single arm). The Josephson current through the system is governed by two control parameters that, in principle, can be varied \textit{in situ}: the relative orientation of the filters magnetization and the magnetic flux through the interferometer loop. We aim at calculating the dependence of the dc Josephson effect on these parameters.

The paper is organized as follows. In Sec.~\ref{sec: model}, we present a detailed description of the considered model. Sections~\ref{sec: andreev_levels} and \ref{sec: josephson_current} contain the obtained results for the Andreev levels and the Josephson current, respectively. Discussion of possible experiments is given in Sec.~\ref{sec:discussion}, and conclusions are presented in Sec.~\ref{sec: conclusion}. Some additional clarifying details are presented in the appendices.

\section{Model}\label{sec: model}

The sketch of Fig.~\ref{fig: model_sketch} implies subdivision of the system into blocks, which can be conveniently described in the framework of the scattering matrix formalism \cite{Beenakker 1991,Nazarov_Blanter}. From left to right the blocks are: S$_1$N interface (gray/blue boundary), left beam splitter (blue), ferromagnetic spin filters (green), right beam splitter (blue), and NS$_2$ interface (blue/gray boundary). The two interferometer arms (upper and lower) are assumed to be identical up to the magnetization direction in the spin filtering parts. In the absence of the superconducting phase difference $\varphi$, the system is also assumed to be left-right symmetric.

In order to elucidate the mechanisms of Josephson transport due to split Cooper pairs in ballistic spin-filtering SQUID, we assume that the beam splitters and the arms are composed of highly transmissive single-channel conductors,
and the ferromagnetic inserts are half-metallic (i.e., providing absolute spin filtering). We also assume zero temperature. Note that interference effects on normal transport in single-channel interferometers have already been studied for a long time \cite{Gefen1984,Datta1985}. Our purpose is to consider supercurrents through similar structures.

We assume the short-junction limit implying that the length of the junction (the distance between the superconductors) is short compared to the coherence length $\xi =\hbar v_F /\Delta$; in this case, the supercurrent is carried by subgap bound states within the junction \cite{Beenakker 1991}. Employing the Bogoliubov--de Gennes (BdG) equations in the framework of the scattering matrix formalism, we follow the standard strategy,
first calculating the Andreev spectrum $E(\varphi)$ and then finding the Josephson current at zero temperature as \cite{Beenakker 1991,Nazarov_Blanter,comment2}
\begin{equation}\label{eq: current-phase_through_spectrum}
I(\varphi)=-\frac{e}{\hbar}\sum_{E\in[0,\Delta]} \frac{dE}{d\varphi}.
\end{equation}

The Andreev levels should be found from the spin-dependent BdG equations \cite{deGennesBook},
\begin{equation}\label{eq: BdG}
H_\mathrm{BdG}
\begin{pmatrix}
u_{\uparrow} \\
u_{\downarrow} \\
v_{\uparrow} \\
v_{\downarrow} \\
\end{pmatrix}
=E
\begin{pmatrix}
u_{\uparrow} \\
u_{\downarrow} \\
v_{\uparrow} \\
v_{\downarrow} \\
\end{pmatrix},
\end{equation}
where the BdG Hamiltonian
\begin{equation} \label{HBdG}
H_\mathrm{BdG} =
\begin{pmatrix}
H_{\uparrow\uparrow} & H_{\uparrow\downarrow} & 0 & \Delta e^{i\varphi} \\
H_{\downarrow\uparrow} & H_{\downarrow\downarrow} & -\Delta e^{i\varphi} & 0 \\
0 & -\Delta e^{-i\varphi} & -H_{\uparrow\uparrow}^{*} & -H_{\uparrow\downarrow}^{*} \\
\Delta e^{-i\varphi} & 0 & -H_{\downarrow\uparrow}^{*} & - H_{\downarrow\downarrow}^{*}
\end{pmatrix}
\end{equation}
acts in the direct product of the particle-hole and spin spaces [$(u,v)$ and $(\uparrow,\downarrow)$ structure of the eigenstates, respectively].
Here, the single-particle Hamiltonian $H_{\alpha\beta}$ (acting in the spin space),
the absolute value of the order parameter $\Delta$, and its phase $\varphi$, depend on spatial coordinates. We will actually consider systems in which magnetism (included into $H_{\alpha\beta}$) and superconductivity (described by $\Delta$ and $\vp$) are spatially separated.

For the system of Fig.~\ref{fig: model_sketch}, the single-particle Hamiltonian has the form
\begin{equation}\label{eq: single electron Hamiltonian}
H_{\alpha\beta}= \left[ \frac{1}{2m}\left(\textbf{p}-\frac{e}{c}\textbf{A}\right)^2-E_F \right] \delta_{\alpha\beta} +U_{\alpha\beta}^\mathrm{(exch)},
\end{equation}
where $E_F$ is Fermi energy and the exchange part $U^\mathrm{(exch)}$ describes the two ferromagnetic arms.

Although the excitation energies $E$ should be positive, it is sometimes convenient to argue in terms of the ``semiconductor model'', in which the negative-energy states also exist and are filled in accordance with the Fermi distribution \cite{Tinkham}. In this respect, it is important that the BdG Hamiltonian (\ref{HBdG}) possesses the particle-hole symmetry
\begin{equation}\label{eq: fundumental symmetry of BdG Hamiltonian}
\lbrace H_\mathrm{BdG}, \mathcal{P}\rbrace=0,
\end{equation}
where $\lbrace\cdot,\cdot\rbrace$ denotes the anticommutator and the symmetry operator is $\mathcal{P}=\sigma_x^\mathrm{PH}\sigma_{0}^\mathrm{S} \mathcal{K}$. Here, $\mathcal{K}$ is the complex conjugation operator, and the Pauli matrices act in the particle-hole (PH) or spin (S) space.
As a consequence, the eigenstates always come in pairs $|\psi\rangle$ and $\mathcal{P} |\psi\rangle$ with mirror-symmetric energies $\pm E$.

Within the scattering matrix formalism, each block of the structure should be described by the corresponding scattering matrix linking an incoming $(u_{\uparrow},u_{\downarrow},v_{\uparrow},v_{\downarrow})^T$ state to the outgoing one \cite{Beenakker 1991,Nazarov_Blanter}. For a scatterer with several terminals labeled by $a,b,\dots$, we define the scattering states as
\begin{equation} \label{eq: statebasis}
(u_{\uparrow a},u_{\downarrow a},u_{\uparrow b},u_{\downarrow b},\dots,v_{\uparrow a},v_{\downarrow a},v_{\uparrow b},v_{\downarrow b},\dots)^T.
\end{equation}
The scattering matrix for an $n$-terminal scatterer has dimensions $4n \times 4n$.

The outmost blocks of our system, Fig.~\ref{fig: model_sketch}, are the $\mathrm{S_1 N}$ and $\mathrm{NS_2}$ interfaces, where subgap quasiparticles inevitably experience the Andreev reflection.
Each interface by itself is described by the $4\times 4$ scattering matrix containing the reflection phases:
\begin{equation}\label{eq: Sigma matrix}
\Sigma_{j}=e^{-i\chi}
\begin{pmatrix}
0 & 0 & 0 & e^{i\varphi_j} \\
0 & 0 & -e^{i\varphi_j} & 0 \\
0 & -e^{-i\varphi_j} & 0 & 0 \\
e^{-i\varphi_j} & 0 & 0 & 0 \\
\end{pmatrix},
\end{equation}
where
\begin{equation} \label{chi}
\chi=\arccos\left(\frac{E}{\Delta}\right)\in\left[0,\pi\right],
\end{equation}
$j=1,2$ indexes the superconducting reservoirs, and $\varphi_{1,2}$ is the phase of the superconducting order parameter in the corresponding reservoir.

When considered together, the two SN interfaces effectively form a two-terminal scatterer. In the basis of Eq.\ (\ref{eq: statebasis}), the corresponding $8\times 8$ scattering matrix has the following form:
\begin{equation}
S_{\Sigma}=
e^{-i\chi}
\begin{pmatrix}
0 & S_{eh} \\
S_{he} & 0
\end{pmatrix},
\end{equation}
where
\begin{equation} \label{eq: SehShe}
S_{eh}=
-S_{he}^{*}=
\begin{pmatrix}
i\sigma_y^{\mathrm{S}} e^{i\varphi_1} & 0 \\
0 & i\sigma_y^{\mathrm{S}} e^{i\varphi_2}
\end{pmatrix}.
\end{equation}

The rest of the structure is formed by nonsuperconducting scatterers, in which $\Delta=0$. Description of these parts is especially convenient in basis (\ref{eq: statebasis}).
First, since electrons do not mix with holes in the absence of superconductivity, the corresponding scattering matrices are block-diagonal in the PH space.
Second, in the short-junction limit, the difference between the wave vectors for electrons and holes can be neglected, hence, the phases accumulated by holes are inverted with respect to the phases accumulated by electrons. As a result, the electron and hole blocks are related simply by complex conjugation:
\begin{equation} \label{SnonS}
S_\mathrm{nonS}=
\begin{pmatrix}
S_e & 0
\\
0 & S_h
\end{pmatrix},
\qquad
S_h = S_e^*.
\end{equation}

We discuss nonsuperconducting elements of our system (the beam splitters and the spin filters) in Secs.~\ref{sec: S-matrices of beam splittres} -- \ref{sec: S-matrix interferometer} below.

\subsection{Beam splitters}\label{sec: S-matrices of beam splittres}

A quasiparticle, Andreev reflected from a superconductor, is divided between the two interferometer arms. We describe this process in terms of three-terminal beam splitters (blue parts in Fig.~\ref{fig: model_sketch}).

Generally, the scattering matrix of the splitter is unitary. In addition, we gauge the magnetic field such that the vector potential exists only inside the spin filters (green regions in Fig.~\ref{fig: model_sketch}) and vanishes elsewhere. Hence the time reversal symmetry is preserved for the beam splitters, and the scattering matrix is symmetric.

We assume geometrical ``Y symmetry'' of each splitter, meaning that one terminal (the one touching a superconductor; number 3) is special, while the two others (connecting to the two interferometer arms; numbers 1 and 2) are equivalent. The core part of the splitter scattering matrix is a symmetric unitary $3\times 3$ matrix $Y$ describing splitting of (spinless) electrons between the three terminals. The Y symmetry then implies two independent relations: $Y_{11} = Y_{22}$ and $Y_{13}=Y_{23}$. In order to focus on processes of Cooper pair splitting, we additionally assume $Y_{33}=0$, meaning that an electron coming from terminal 3 is not reflected back but only transmitted to terminals 1 and 2.

The most general form of the $3\times 3$ scattering matrix satisfying the above restrictions is parametrized by two real phases, $\alpha$ and $\beta$, as
\begin{equation} \label{Y}
Y=
\begin{pmatrix}
-\frac{e^{i\alpha}}2 & \frac{e^{i\alpha}}2 & \frac{e^{i(\alpha/2+\beta)}}{\sqrt{2}} \\
\frac{e^{i\alpha}}2 & -\frac{e^{i\alpha}}2 & \frac{e^{i(\alpha/2+\beta)}}{\sqrt{2}} \\
\frac{e^{i(\alpha/2+\beta)}}{\sqrt{2}} & \frac{e^{i(\alpha/2+\beta)}}{\sqrt{2}} & 0
\end{pmatrix}.
\end{equation}
Physically, $\alpha/2$ is the phase accumulated by electrons moving along the ``legs'' of terminals 1 or 2 (see Fig.~\ref{fig: model_sketch}), while $\beta$ is the corresponding phase for terminal 3.

The $12\times 12$ scattering matrix of the beam splitter is an extension of the $Y$ matrix to include the S and PH degrees of freedom. The PH structure corresponds to the form determined by Eq.\ (\ref{SnonS}). As there is no exchange field in the splitters, the matrix is trivial in the spin space, with the electron block given by
\begin{equation} \label{SBS}
S_\mathrm{BS} =
Y \sigma_0^\mathrm{S}
\end{equation}

All closed quasiparticle trajectories (responsible for the formation of Andreev levels) include only multiples of $2\times \alpha/2=\alpha$, therefore all physically different values of $\alpha$ lie in the range $\left[0,2\pi\right]$.

Since any subgap electron falling from leg 3 onto the SN interface is inevitably Andreev reflected as a hole, and the phases accumulated by the electron and the hole in leg 3 compensate each other, the corresponding phase $\beta$ does not influence our results, dropping out from calculations below.
Therefore, without loss of generality, we put $\beta=0$.

\subsection{Spin filters}\label{sec: S-matrix of spin filters}
To model the effect of spin filtering by ferromagnets (green insets in Fig.~\ref{fig: model_sketch}), we assume ideal spin filters (the limit of half-metallic ferromagnets): an incoming quasiparticle with spin $\mathbf s$ along the direction $\mathbf n$ of the filter magnetization is transmitted without any reflection. In the opposite case, the filter acts as the infinitely high barrier for the quasiparticle.
In terms of the exchange energy, this implies that
\begin{equation}
U^\mathrm{(exch)}=\left\{
\begin{array}{cc}
0, & \;\bm{s}\uparrow\uparrow \bm{n}, \\
\infty, & \;\bm{s}\uparrow\downarrow \bm{n}. \\
\end{array}\right.
\end{equation}

Each filter is described by $8\times 8$ scattering matrix with the PH structure determined by Eq.\ (\ref{SnonS}).
The $4\times 4$ electron blocks has the form
\begin{equation} \label{SF}
S_\mathrm{F}=
\begin{pmatrix}
r_{11} & t_{12} \\
t_{21} & r_{22}
\end{pmatrix}.
\end{equation}
The reflection blocks in our model are
\begin{equation}
r_{11} = r_{22} = R_{\bm{n}}
\begin{pmatrix}
0 & 0 \\
0 & -1
\end{pmatrix}
R^{-1}_{\bm{n}},
\end{equation}
where the $R_{\bm{n}}$ matrix takes into account the allowed spin direction inside the filter, $\bm{n} = (\sin\Omega \cos\phi, \sin\Omega \sin\phi, \cos\Omega)$:
\begin{equation}\label{eq: spinor rotation matrix}
R_{\bm{n}}=
\begin{pmatrix}
\cos\frac{\Omega}{2} & -e^{-i\phi}\sin\frac{\Omega}{2} \\
e^{i\phi}\sin\frac{\Omega}{2} & \cos\frac{\Omega}{2} \\
\end{pmatrix}.
\end{equation}
In the geometry of Fig.~\ref{fig: model_sketch}, we have $\phi=0$, while $\Omega = \theta/2$ and $\Omega=-\theta/2$ in the upper and lower arms, respectively.

The transmission blocks of $S_\mathrm{F}$ for the upper arm have the form
\begin{equation} \label{t}
t_{12}(\Phi)=t_{21}(-\Phi)
=R_{\bm{n}}
\begin{pmatrix}
e^{i\left(L+\pi f/2\right)} & 0 \\
0 & 0 \\
\end{pmatrix}
R^{-1}_{\bm{n}}.
\end{equation}
Here, $L$ is the geometric phase (proportional to the length of the filter) and $\pi f/2$ is the magnetic phase, with $f=\Phi /\Phi_0$ being the dimensionless external flux (in our gauge, the vector potential is nonzero only inside the ferromagnetic insets, where it equals $\Phi /2L$). For the lower arm, the sign of the magnetic phase is inverted.

Physically, the above matrices imply that depending on its spin, a quasiparticle either passes through the filter accumulating phase $\pm(L\pm \pi f/2)$ (the signs correspond to electons/holes and upper/lower arm) or is reflected from infinitely high barrier with phase $\pi$.

Note that in our model, we have introduced two geometrical phases, $\alpha$ and $L$, determined by the lengths of the segments that can belong to the same arm of the interferometer loop. This may seem redundant; however, their role is different due to nontrivial interference effects caused by them. For instance, if try to get rid of $\alpha$ putting $\alpha=0$, the interferometer loop effectively becomes completely opaque due to destructive interference \cite{Gefen1984} (in the special case of $\theta=\pi$, this effect has a simple explanation, see Appendix~\ref{App: interferential opacity of splitters}). At the same time, at $L=0$, the behavior of the system loses many characteristic features. We therefore keep both the phases as geometrical parameters of our system.

\subsection{Scattering matrix of the interferometer}\label{sec: S-matrix interferometer}

Combining the scattering matrices of the two spin filters and the two beam splitters, we calculate the $8\times 8$ scattering matrix of the whole interferometer loop:
\begin{equation}\label{eq: S_interferometer}
S_\mathrm{interf}=
\begin{pmatrix}
S_{ee} & 0 \\
0 & S_{hh}
\end{pmatrix}.
\end{equation}
The electron and hole blocks have the following structure:
\begin{equation} \label{eq: See and Shh}
S_{ee}=S_{hh}^{*}=
\begin{pmatrix}
r_{\uparrow} & r & t_{\uparrow} & t \\
-r & r_{\downarrow} & t & t_{\downarrow} \\
t_{\uparrow} & -t & r_{\uparrow} & -r \\
-t & t_{\downarrow} & r & r_{\downarrow}
\end{pmatrix},
\end{equation}
with the following symmetry relations for the elements:
\begin{gather}
r(-\Phi)=-r(\Phi),
\quad
r_{\uparrow,\downarrow}(-\Phi)=r_{\uparrow,\downarrow}(\Phi),
\notag \\
t(-\Phi)=-t(\Phi),
\quad
t_{\uparrow,\downarrow}(-\Phi)=t_{\uparrow,\downarrow}(\Phi).
\end{gather}

The matrix elements in Eq.\ (\ref{eq: See and Shh}) are straightforwardly obtained within our formalism; however, their explicit form is cumbersome, and we do not provide the corresponding expressions here.

\section{Andreev levels}\label{sec: andreev_levels}

Having found the scattering matrix of the interferometer, we reduced the problem to a very general formulation, in which a nonsuperconducting [i.e., diagonal in the PH space, see Eq.\ (\ref{eq: S_interferometer})] scatterer provides Josephson coupling between two superconductors. In the short-junction limit, when
the energy dependence of the scattering matrix can be neglected,
we can obtain explicit analytical expressions for the Andreev levels inside the junction for arbitrary nonsuperconducting scatterer.

\subsection{General analytical expression}

In this section, we find the Andreev levels in the case of arbitrary nonsuperconducting scatterer between the superconductors in the short-junction limit.
We still use notation of Eq.\ (\ref{eq: S_interferometer});
however, the calculation below is not specific for our setup, and requires only the diagonal PH structure of the scattering matrix and the short-junction limit. In particular, the scattering matrix can have arbitrary spin structure (due to ferromagnetism, spin-orbit interaction, etc.).

The standard procedure requiring existence of nontrivial solution in the matching problem for the scattering states (in other words, existence of an eigenstate of $H_\mathrm{BdG}$) \cite{Beenakker 1991,Nazarov_Blanter}, yields the spectral equation
\begin{equation}\label{eq: non_simplified_spectral_equation}
\det\left[\widehat{1}_{8\times 8}-
e^{-i\chi}
\begin{pmatrix}
0 & S_{eh} \\
S_{he} & 0
\end{pmatrix}
\begin{pmatrix}
S_{ee} & 0 \\
0 & S_{hh}
\end{pmatrix}
\right]
=0,
\end{equation}
with $S_{eh}$ and $S_{he}$ given by Eq.\ (\ref{eq: SehShe}).

Taking the determinant in the PH space,
we can rewrite the spectral equation as
\begin{equation} \label{det4x4}
\det\left[e^{2i\chi}\cdot\widehat{1}_{4\times 4}-S_{eh} S_{hh}S_{he}S_{ee}\right]=0.
\end{equation}

In this equation, energy $E$ of Andreev levels is encoded in $\chi$ [see Eq.\ (\ref{chi})].
Denoting $\lambda = e^{2i\chi}$, we see that in order to find the Andreev levels, we need to calculate the eigenvalues $\lambda_{1,\dots,4}$ of matrix $M=S_{eh} S_{hh}S_{he}S_{ee}$ (at this point, it is important that the short-junction limit is assumed so that $M$ does not depend on $E$).

Due to the BdG symmetry (\ref{eq: fundumental symmetry of BdG Hamiltonian}), the Andreev levels always come in pairs $\pm E$, hence the eigenvalues of $M$ come in conjugate pairs:
\begin{equation}
\lambda_1=\Lambda_1, \quad \lambda_2=\Lambda_1^{*}, \quad \lambda_3=\Lambda_2, \quad \lambda_4=\Lambda_2^{*},
\end{equation}
where $\vert \Lambda_{1,2}\vert=1$ [this structure of the eigenvalues can be explicitly seen from relations (\ref{eq: SehShe}) and (\ref{SnonS}), implying that $M$ is a special unitary matrix with $\det M=1$].

The eigenvalues $\Lambda_{1,2}$ can be found with the help of the following trick.
On one hand, straightforward calculation of the determinant yields
\begin{equation}\label{eq: second form of det lambda-M}
\det\left[\lambda\cdot\widehat{1}_{4\times 4}-M\right]=\lambda^4-\lambda^3\cdot \tr M+\dots +1,
\end{equation}
while, on the other hand, in terms of the eigenvalues we can write
\begin{gather}
\det\left[\lambda\cdot\widehat{1}_{4\times 4}-M\right] =\left(\lambda-\Lambda_1\right)\left(\lambda-\Lambda_1^{*}\right)\left(\lambda-\Lambda_2\right)\left(\lambda-\Lambda_2^{*}\right)
\notag \\
=\lambda^4-\lambda^3\cdot \left(2\Re\Lambda_1+2\Re\Lambda_2\right)+\dots+1.
\label{eq: first form of det lambda-M}
\end{gather}

Comparing the coefficients in front of $\lambda^3$ and denoting
$\mathcal T= \tr M$,
we obtain the first equation for $\Lambda_{1,2}$:
\begin{equation}\label{eq: first eq for ReLamda}
\mathcal T=2\Re\Lambda_1+2\Re\Lambda_2.
\end{equation}
The second equation can be obtained from Eq.\ (\ref{eq: first form of det lambda-M}) if we consider $\lambda$ as a free variable and put, for example, $\lambda =i$. Then, denoting $\mathcal D=\det\left[i\cdot\widehat{1}_{4\times 4}-M \right]$, we find
\begin{equation}\label{eq: second eq for ReLamda}
\mathcal D=-4\Re\Lambda_1\cdot \Re\Lambda_2.
\end{equation}

Expressing $\Re\Lambda_{1,2}$ from Eqs.\ (\ref{eq: first eq for ReLamda}) and (\ref{eq: second eq for ReLamda}) and recalling that $\Re\Lambda=\Re e^{2i\chi} = 2 ( E/\Delta )^2-1$, we find analytical expression for two positive Andreev levels,
\begin{equation}\label{eq: analytical expression for spectrum}
E= \frac{\Delta}{2\sqrt{2}}\left[\mathcal T+4\pm\sqrt{\mathcal T^2+4\mathcal D}\right]^{1/2},
\end{equation}
while the two negative levels only differ by the sign. In total, as a consequence of the short-junction limit, we have only four Andreev levels (while in the general case, their number can be larger and grows with the junction length).

Equation (\ref{eq: analytical expression for spectrum}) is a general expression for the Andreev levels in the case of arbitrary nonsuperconducting scatterer in the short-junction limit. In order to show its relation to known results, we can consider the special case of a spin-independent scattering matrix,
\begin{equation} \label{eq: See}
S_{ee}=\begin{pmatrix}
r_{ee} & t_{ee}' \\
t_{ee} & r_{ee}'
\end{pmatrix}
\sigma_0^\mathrm{S}.
\end{equation}
This results in $\mathcal T^2 + 4\mathcal D = 0$, so that the square root in Eq.\ (\ref{eq: analytical expression for spectrum}) vanishes, leading to double degeneracy of the levels. Considering the general case of possibly broken time-reversal symmetry, such that $t_{ee}' = t_{ee} e^{i\delta}$ with nonzero $\delta$, we find
\begin{equation}
\mathcal T = 4 \left( 1- 2 T \sin^2 \frac{\varphi+\delta}2 \right),
\end{equation}
with $T=| t_{ee} |^2$ being the transparency of the channel. As a result, Eq.\ (\ref{eq: analytical expression for spectrum}) yields
\begin{equation} \label{eq: Beenakker_with_delta}
E=\Delta\left[1-T \sin^2\frac{\varphi+\delta}{2}\right]^{1/2},
\end{equation}
which in the time-reversal-symmetric case ($\delta=0$) reproduces the well-known result for the quantum point contact (QPC) \cite{Beenakker 1991}.

Below, we apply the general analytical expression (\ref{eq: analytical expression for spectrum}) for the Andreev spectrum in short Josephson junctions to the SQUID system of Fig.~\ref{fig: model_sketch}.

\subsection{Special cases} \label{sec:specialcases}

\begin{figure*}
\centering{\includegraphics[width=\textwidth]{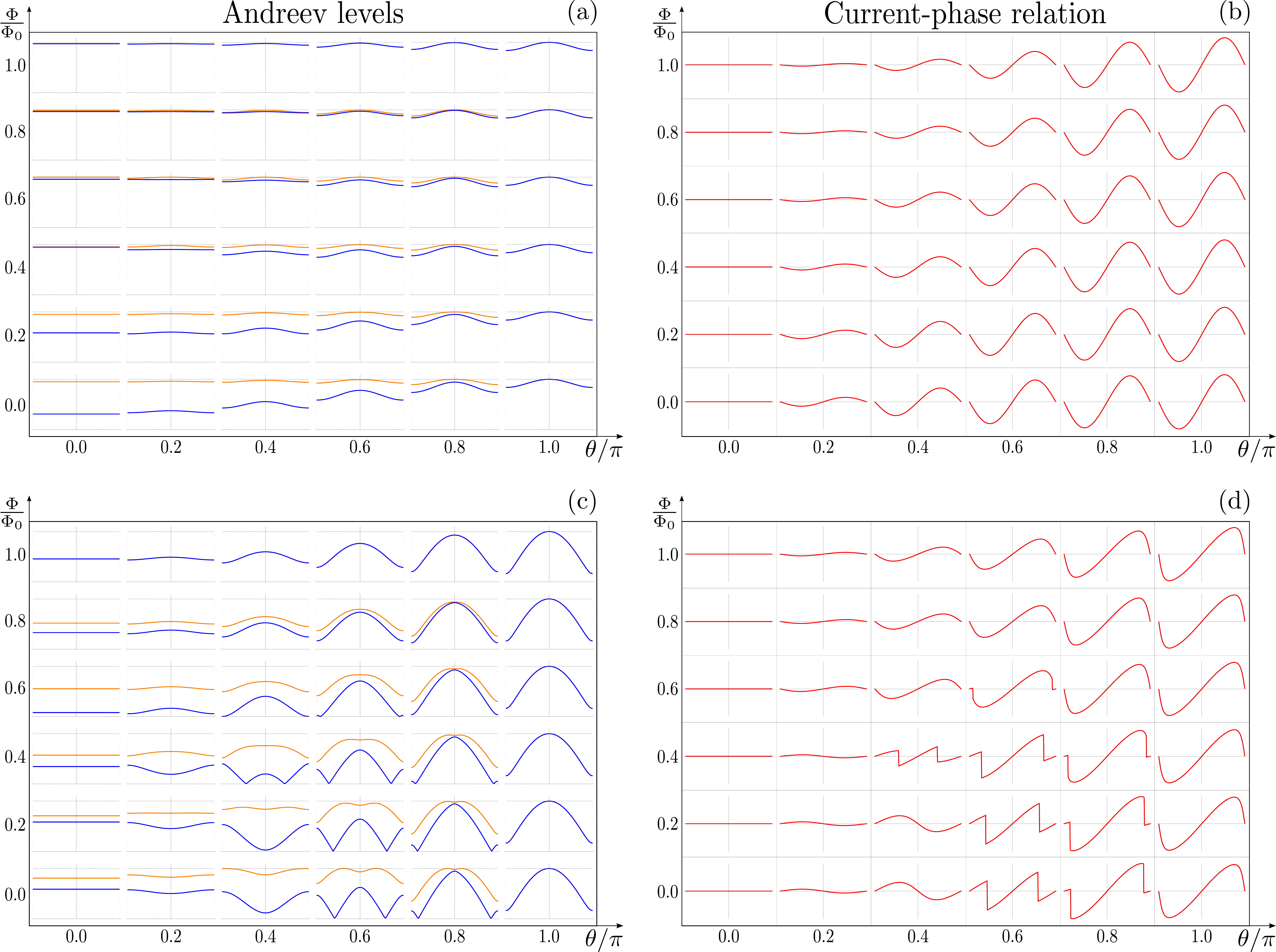}}
\caption{
Dependence of the Andreev spectrum [(a) and (c)] and the current-phase relation [(b) and (d)] on angle $\theta$ between the magnetizations of the filters and on the magnetic flux $\Phi$ for two different sets of geometrical parameters (upper and lower rows). Each panel is a collection of many tiles, each built at the values of $\theta$ and $\Phi$ corresponding to the position of the tile. In each tile, the horizontal axis corresponds to the phase difference $\varphi\in\left[-\pi,\pi\right]$. The vertical axes correspond to the energy ($E\in\left[0,\Delta\right]$) and the supercurrent $I$ (measured in units of maximal critical current $I_c^{\mathrm{max}}$ for the corresponding geometry) for the left and right columns, respectively. The dependence on $\Phi$ is $2\Phi_0$-periodic. Only two positive Andreev levels are shown (the two negative levels differ by sign). For some geometries, the behavior is simple [(a) and (b), corresponding to $\alpha=0.25\pi$ and $L=0.8\pi$]; as $\theta$ decreases, the Andreev levels become flatter, and the critical current decreases, reaching zero for parallel magnetizations.
For some geometries, the behavior is much more complicated [(c) and (d), $\alpha=0.5\pi$, $L=0.4\pi$]; as $\theta$ decreases, level crossings at zero energy may appear, leading to discontinuities in the current-phase characteristics. For some values of $\theta$, the system demonstrates a $\pi$-junction-type current-phase relation [see, e.g., $\theta=0.4\pi$, $\Phi=0$ in (d)].}
\label{fig: andreev_levels_current_phase}
\end{figure*}

Dependence of the Andreev spectrum
on the interferometer control parameters $\theta$ and $\Phi$ is shown in Figs.~\ref{fig: andreev_levels_current_phase}(a) and \ref{fig: andreev_levels_current_phase}(c) for two different sets of geometrical parameters $\alpha$ and $L$.
Keeping Figs.~\ref{fig: andreev_levels_current_phase}(a) and \ref{fig: andreev_levels_current_phase}(c) in mind, we first present and discuss analytical results for the spectrum in special cases in Secs.~\ref{sec: parallel_magnetizations} -- \ref{sec: half_period_flux}. Numerical results for the general case are then discussed in Sec.~\ref{sec: general_case}.

\subsubsection{Parallel magnetizations, $\theta=0$}\label{sec: parallel_magnetizations}
When the magnetizations of the two filters are parallel, transport of split Cooper pairs is completely suppressed, and the supercurrent is zero.
Indeed, transfer of a Cooper pair across the system from S$_1$ to S$_2$ is equivalent in the BdG language to transfer of an electron from S$_1$ to S$_2$, its Andreev reflection into a hole and subsequent transfer of the hole back from S$_2$ to S$_1$. This process is impossible at $\theta=0$: if the electron has passed through a spin filter, this means that its spin is along the allowed direction of both the filters, hence the Andreev reflected hole with opposite spin cannot travel back to S$_1$, being trapped between S$_2$ and the filters.

\begin{figure}[t]
\includegraphics[width=\columnwidth]{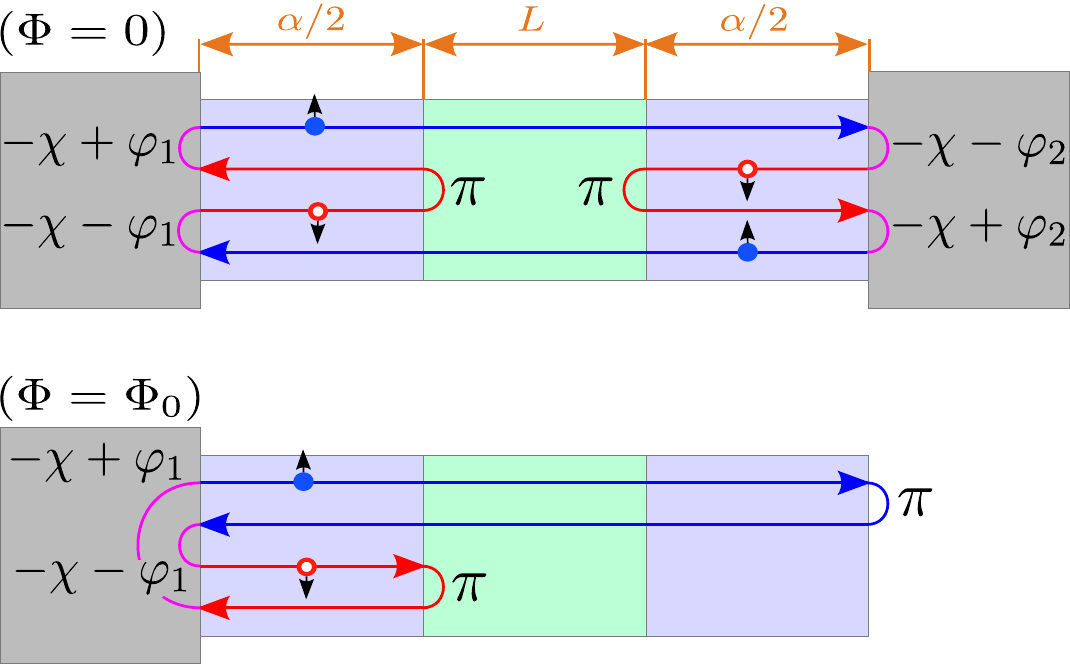}
\caption{
Quasiparticle trajectories in the case of parallel spin filters and the magnetic flux 0 or $\Phi_0$. The blue and red lines are electrons and holes, respectively. At $\Phi=0$, the interferometer is perfectly transparent for up spins and completely opaque for down spins [see Eq.\ (\ref{eq: S_ee at zero theta and zero Phi})]. As phase $-4\chi+2L$, accumulated  along closed trajectory, must be a multiple of $2\pi$, the corresponding energy levels are given by Eq.\ (\ref{eq: parllel_magnetizations_zero_flux}). At $\Phi=\Phi_0$, the loop is effectively opaque for up spins as well [beams from the upper and lower arms interfere destructively, see Eq.\ (\ref{eq: S_ee at zero theta and Phi_0 Phi})]. The phase along the loop is $-2\chi+2L+\alpha$, and the level is given by Eq.\ (\ref{eq: parllel_magnetizations_phi_flux}).
}
\label{fig: parallel_magnetizations_trajectories}
\end{figure}

The trajectory can finally be closed after another Andreev reflection, turning the hole into electron near S$_2$ (see Fig.~\ref{fig: parallel_magnetizations_trajectories}). However, this process and, hence, the energy levels are insensitive to the phase difference $\varphi$, and depend only on the magnetic flux $\Phi$ and the geometrical parameters $\alpha$ and $L$. Such levels do not carry supercurrent; still, it is instructive to consider them in order to get some insight into physical processes specific for our system.

From Eq.\ (\ref{eq: analytical expression for spectrum}), we obtain
\begin{widetext}
\begin{equation}\label{eq: parallel_magnetization_spectrum_general_expression}
E=\frac{\Delta}{\sqrt{2}}\left[
1+
\frac{\sin^2 \left(\frac{\pi}{2}f\right) \left[ \cos(\alpha+2L)-\cos\alpha \cos^2\left( \frac{\pi}{2}f \right)\right]
\pm 2 \cos\left(\frac{\pi}{2}f\right) \sin(\alpha+L) \left[ \sin\alpha \cos^2 \left(\frac{\pi}{2}f\right) + \sin(\alpha+2L) \right]}
{\sin^4 \left(\frac{\pi}{2}f\right) + 4 \cos^2 \left(\frac{\pi}{2}f \right)\sin^2 (\alpha +L)}
\right]^{1/2}
.
\end{equation}
\end{widetext}

In the case of $\Phi=0$, the expression (\ref{eq: parallel_magnetization_spectrum_general_expression}) simplifies as
\begin{equation}\label{eq: parllel_magnetizations_zero_flux}
E =\Delta\left| \cos (L/2) \right|,\quad E = \Delta\left| \sin (L/2) \right|,
\end{equation}
while at $\Phi=\Phi_0$ (i.e., at $f=1$) we find a doubly degenerate level
\begin{equation}\label{eq: parllel_magnetizations_phi_flux}
E = \Delta \left|\cos\left(L+\alpha/2\right)\right|.
\end{equation}

In both the cases, the scattering matrix of the interferometer simplifies considerably. Namely,
at $\Phi=0$, the elements (\ref{eq: See and Shh}) of the scattering matrix are as follows:
\begin{equation}
\begin{aligned}\label{eq: S_ee at zero theta and zero Phi}
r_{\uparrow}&=0, & r_{\downarrow}&=-e^{i\alpha}, & r&=0, \\
t_{\uparrow}&=e^{i(\alpha+L)}, & t_{\downarrow}&=0, & t&=0,
\end{aligned}
\end{equation}
while at $\Phi=\Phi_0$, we obtain
\begin{equation}\label{eq: S_ee at zero theta and Phi_0 Phi}
\begin{aligned}
r_{\uparrow}&=-e^{2i(L+\alpha)}, & r_{\downarrow}&=-e^{i\alpha}, & r&=0, \\
t_{\uparrow}&=0, & t_{\downarrow}&=0, & t&=0.
\end{aligned}
\end{equation}

As a result, energy levels (\ref{eq: parllel_magnetizations_zero_flux}) and (\ref{eq: parllel_magnetizations_phi_flux}) can be understood by calculating the phase along closed quasiparticle trajectories (see Fig.~\ref{fig: parallel_magnetizations_trajectories}).

\subsubsection{Antiparallel magnetizations, $\theta=\pi$}\label{sec: antiparallel_magnetizations}

When the two spin filters are magnetized in opposite directions ($\theta=\pi$), the scattering matrix of the interferometer, Eq.\ (\ref{eq: See and Shh}), has the following simplified structure:
\begin{equation} \label{eq: S_ee theta=pi and arbitrary flux}
S_{ee}=
\begin{pmatrix}
r_0 \sigma_0^\mathrm{S}  &  t_0 \exp\left(-i\sigma_x^\mathrm{S} \frac{\pi f}{2}\right)
\\
t_0 \exp\left( i\sigma_x^\mathrm{S} \frac{\pi f}{2} \right) & r_0 \sigma_0^\mathrm{S}
\end{pmatrix},
\end{equation}
where $r_0$ and $t_0$ are complex reflection and transmission amplitudes.

The spin structure of Eq.\ (\ref{eq: S_ee theta=pi and arbitrary flux}) implies that reflection is trivial in spin,
while transmission rotates it by flux-dependent angle $\pi f$ around the $x$ axis (which is collinear to the magnetization directions at $\theta=\pi$).

As a consequence of this rotation, the Andreev bound states are ``spin-twisted''. 
An electron with spin-up near S$_1$ is transformed into electron rotated by the angle $\pi f$ near S$_2$ (each electron is accompanied by the corresponding Andreev-reflected hole). In the limiting case of zero magnetic flux, $\Phi=0$, the rotation disappears, while at $\Phi=\Phi_0$, the rotation angle is equal to $\pi$, hence the spins near the left and right superconductors (S$_1$ and S$_2$) are flipped with respect to each other.

Although the magnetic flux strongly modifies the structure of the Andreev bound states, it does not influence the spectrum. Indeed, the rotation angle $\pi f$ drops out from the product $M=S_{eh} S_{hh}S_{he}S_{ee}$ [see Eq.\ \ref{det4x4}], thus the magnetic flux does not enter the final expression (\ref{eq: analytical expression for spectrum}) for the spectrum. 
The physical reason is that an electron, rotated after passing the interferometer, is then Andreev-reflected, and when traveling back, the hole 
undergoes compensating rotation.
Consequently, for calculating the spectrum, we may put $\Phi=0$ for simplicity.

Next, at $\Phi=0$, the angle $\pi f$ vanishes and the scattering matrix of the interferometer (\ref{eq: S_ee theta=pi and arbitrary flux}) becomes trivial in spin. The time-reversal symmetry is therefore effectively restored and spectrum of our system should reproduce the well-known result for the quantum point contact (QPC) Josephson junction \cite{Beenakker 1991} [similarly to Eqs.\ (\ref{eq: See})--(\ref{eq: Beenakker_with_delta}) with $\delta=0$]. Indeed, from Eq.\ (\ref{eq: analytical expression for spectrum}), we obtain
\begin{equation}\label{eq: spectrum of short junction}
E=\Delta\left[1-T_0\sin^2\frac{\varphi}{2}\right]^{1/2},
\end{equation}
with effective transparency
$T_0 = |t_0|^2$
determined by the geometrical parameters of our system:
\begin{equation} \label{T0}
T_0= \frac{16 \sin^4 \frac\alpha 2}
{16 \sin^4 \frac\alpha 2 +4 \sin^2 \frac\alpha 2 +\sin^2 L +4 \sin\left( 2L +\frac{3\alpha}2 \right) \sin\frac\alpha 2}.
\end{equation}

\subsubsection{Zero external flux, $\Phi=0$}\label{sec: zero_flux}
The interferometer scattering matrix is nontrivial in spin. At the same time, since the filters' magnetizations are symmetric with respect to the $z$ axis (see Fig.~\ref{fig: model_sketch}), and at $\Phi=0$ the two arms are equivalent from the orbital point of view, the $z$ component of spin is conserved. Indeed, the scattering matrix of the interferometer, Eq.\ (\ref{eq: See and Shh}), takes the following form:
\begin{equation}\label{eq: S-matrix for zero external flux}
S_{ee}=e^{i A}
\begin{pmatrix}
\rho_{\uparrow} & 0 & i \tau_{\uparrow} & 0 \\
0 & \rho_{\downarrow} e^{i\Psi} & 0 & i \tau_{\downarrow} e^{i\Psi} \\
i \tau_{\uparrow} & 0 & \rho_{\uparrow} & 0 \\
0 & i \tau_{\downarrow} e^{i\Psi} & 0 & \rho_{\downarrow} e^{i\Psi}
\end{pmatrix}.
\end{equation}
Here, $\rho_{\uparrow, \downarrow}$, $\tau_{\uparrow,\downarrow}$, $\Psi$, and $A$ are real quantities, which depend on the system parameters $\theta$, $\alpha$, and $L$.

Coefficients $\rho_{\uparrow, \downarrow}$ and $\tau_{\uparrow,\downarrow}$ describe reflection and transmission
(we can choose $\rho_{\uparrow}$ and $\rho_\downarrow$ to be positive, then the signs $\tau_\uparrow$ and $\tau_\downarrow$ can be arbitrary).
Phase $\Psi$ is the difference between the phases accumulated by electron with spin up and down, when passing through the interferometer (the reflection amplitude has the same phase as the corresponding transmission amplitude due to unitarity). Phase $A$ is a common phase of all the matrix elements and it does not enter the final expression for the spectrum. Explicit expressions for these parameters in terms of $\theta$, $\alpha$, and $L$ are extremely cumbersome.

However, the general expression (\ref{eq: analytical expression for spectrum}) for the Andreev levels can still be written in terms of the new parameters:
\begin{widetext}
\begin{equation} \label{EPhi=0}
E=\frac{\Delta}{\sqrt{2}}
\left[1+\left(\rho_{\uparrow} \rho_{\downarrow}+\tau_{\uparrow} \tau_{\downarrow}\cos\varphi\right)\cos\Psi\pm
\sin\Psi \sqrt{1-\left(\rho_{\uparrow} \rho_{\downarrow}+\tau_{\uparrow} \tau_{\downarrow}\cos\varphi\right)^2}\right]^{1/2}.
\end{equation}
\end{widetext}

Note that while this result is valid at $\Phi=0$ and arbitrary angle $\theta$ between magnetizations, the case of parallel magnetizations ($\theta=0$) also falls into the scope of Sec.~\ref{sec: parallel_magnetizations}, therefore Eq.\ (\ref{EPhi=0}) should reproduce Eq.\ (\ref{eq: parllel_magnetizations_zero_flux}) in this limit. Indeed, at $\theta=0$ in Eq.\ (\ref{eq: S-matrix for zero external flux}), we obtain $\rho_{\uparrow}=\tau_{\downarrow}=0$, $\rho_{\downarrow}=\tau_{\uparrow}=1$, $\Psi=-L-\pi/2$, and $A=\alpha+L-\pi/2$, thus reproducing the scattering matrix given by Eqs.\ (\ref{eq: See and Shh}) and (\ref{eq: S_ee at zero theta and zero Phi}). Consequently, Eq.\  (\ref{EPhi=0}) reproduces Eq.\ (\ref{eq: parllel_magnetizations_zero_flux}).

Similarly, in the case of $\Phi=0$ and antiparallel magnetizations ($\theta=\pi$), the result of Eq.\ (\ref{EPhi=0}) should be consistent with Eqs.\ (\ref{eq: spectrum of short junction}) and (\ref{T0}) of Sec.~\ref{sec: antiparallel_magnetizations}. In this limit, the parameters determining the scattering matrix (\ref{eq: S-matrix for zero external flux}) take the following values: $\rho_{\uparrow}=\rho_{\downarrow}=\sqrt{1-T_0}$, $\tau_{\uparrow}=\tau_{\downarrow}=\sqrt{T_0}$, and $\Psi=0$ (the overall phase $A$ is nontrivial but drops out from physical quantities, and we omit it for brevity). As a result, Eq.\ (\ref{EPhi=0}) reproduces Eqs.\ (\ref{eq: spectrum of short junction}) and (\ref{T0}).

In Appendix~\ref{sec: qualitative_explanation_of_spectrum}, we demonstrate how some of regimes shown in Fig.~\ref{fig: andreev_levels_current_phase} for $\Phi=0$, can be explained qualitatively.

\subsubsection{Half-period external flux, $\Phi=\Phi_0$}\label{sec: half_period_flux}

At $\Phi=\Phi_0$, the scattering matrix of the interferometer, Eq.\ (\ref{eq: See and Shh}), takes the following form:
\begin{align}
&S_{ee} =
e^{iB}
\notag \\
& \!\! \times \!\!
\begin{pmatrix}
\sqrt{1-\tau^2}e^{i\psi} & 0 & 0 & \tau \\
0 & \sqrt{1-\tau^2}e^{-i\psi} & \tau & 0 \\
0 & -\tau & \sqrt{1-\tau^2}e^{i\psi} & 0 \\
-\tau & 0 & 0 & \sqrt{1-\tau^2}e^{-i\psi}
\end{pmatrix}.
\end{align}
Here, $\tau$, $\psi$, and $B$ are real: $\tau$ describes transparency of the system, $2\psi$ is the difference between the phases
acquired during down-spin and up-spin reflection, and $B$ is the overall phase which does not enter the final expression for the spectrum.

Interestingly, reflection still conserves the $z$ component of spin, while transmission flips it. The eigenstates are consequently ``spin-flipped'' (in a nonlocal sense): the sector with `electron up near S$_1$ and down near S$_2$ plus the Andreev-reflected holes' and the sector with opposite spins (in the limiting case of $\theta=\pi$, this has already been discussed in Sec.~\ref{sec: antiparallel_magnetizations} at $\Phi =\Phi_0$).
This flipping effectively restores the spin symmetry: the spin-flipped sectors are equivalent (even in the presence of the preferred $z$ direction of the filters configuration), hence the levels (\ref{eq: analytical expression for spectrum}) become degenerate.

Unlike the case of $\theta=\pi$, the time reversal symmetry is now generally broken (i.e., $S_\mathrm{interf} \neq S_\mathrm{interf}^T$) and the system does not straightforwardly reduce to the QPC Josephson junction described by Eq.\ (\ref{eq: spectrum of short junction}). Instead, by simplifying Eq.\ (\ref{eq: analytical expression for spectrum}), we obtain
\begin{equation}
E=\Delta\left[\cos^2\psi+\tau^2\sin^2\psi-\tau^2\sin^2\frac{\varphi}{2}\right]^{1/2}.
\end{equation}

At the same time, this result can still be reduced to the QPC-like form as
\begin{equation} \label{Eeff}
E=\Delta_\mathrm{eff}\left[1-T_\mathrm{eff}\sin^2\frac{\varphi}{2}\right]^{1/2},
\end{equation}
with
\begin{align}
\Delta_{\mathrm{eff}} &= \Delta\sqrt{\cos^2\psi+\tau^2\sin^2\psi},
\label{Deltaeff} \\
T_{\mathrm{eff}} &= \frac{\tau^2}{\cos^2\psi+\tau^2\sin^2\psi}.
\label{Teff}
\end{align}
In terms of the original system parameters, we find
\begin{widetext}
\begin{equation}
\Delta_\mathrm{eff}= \Delta\left[
\frac{16 \sin^4 \frac\alpha 2 \sin^2 \frac\theta 2 +\left[ 1+\cos^2 \frac\theta 2 \right]^2 \sin^2 L -4 \left[ 1+\cos^2 \frac\theta 2 \right] \sin(\alpha+L) \sin L +4 \sin^2(\alpha+L)}
{8 \sin^4 \frac\alpha 2 + 10 \sin^2 \frac\alpha 2 +2\sin \frac\alpha 2 \sin\left( \frac{3\alpha}2+2L \right) + \sin^4 \frac\theta 2 \sin^2 L - 4 \cos\theta \sin L \sin\frac\alpha 2 \cos\left( \frac{3\alpha}2+L \right)}
\right]^{1/2},
\end{equation}
\begin{equation}
T_\mathrm{eff}= \frac{16 \sin^4 \frac\alpha 2 \sin^2 \frac\theta 2}
{16 \sin^4 \frac\alpha 2 \sin^2 \frac\theta 2 +\left[ 1+\cos^2 \frac\theta 2 \right]^2 \sin^2 L -4 \left[ 1+\cos^2 \frac\theta 2 \right] \sin(\alpha+L) \sin L +4 \sin^2(\alpha+L)}.
\end{equation}
\end{widetext}

Nonideal transparency $T_\mathrm{eff}$, generated due to interference effects, is not surprising. At the same time, it is interesting that $T_\mathrm{eff}$ can be tuned by varying $\theta$, i.e., misorientation of the spin filters. On the other hand, effective suppression of $\Delta_\mathrm{eff}$ in the expression for the Andreev levels, Eq.\ (\ref{Eeff}), is rather unexpected since the actual order parameter $\Delta$ of the superconducting reservoirs is not altered.

It is straightforward to check that at $\theta=0$ and $\theta=\pi$, the above expressions agree with the corresponding results of Sec.~\ref{sec: parallel_magnetizations} for parallel and Sec.~\ref{sec: antiparallel_magnetizations} for antiparallel magnetizations.

Indeed, at $\theta=0$, we obtain $T_\mathrm{eff}=0$ and $\Delta_\mathrm{eff}=\Delta |\cos(L+\alpha/2) |$, thus reproducing Eq.\ (\ref{eq: parllel_magnetizations_phi_flux}). Concerning the structure of the Andreev states, the limiting cases of spin conserving (at $\theta=0$) and spin flipping (at $\Phi=\Phi_0$) transmission agree with each other due to the fact that $\tau\vert_{\theta=0}=0$, and the interferometer becomes opaque (i.e., transmission disappears).

In the case of $\theta=\pi$, we obtain $T_\mathrm{eff}=T_0$ and $\Delta_\mathrm{eff}=\Delta$, thus reproducing Eqs.\ (\ref{eq: spectrum of short junction}) and (\ref{T0}). Half-period flux $\Phi=\Phi_0$ implies $f=1$, thus spin twisting in Eq.\ (\ref{eq: S_ee theta=pi and arbitrary flux}) reduces to spin flipping (i.e., $\pi$ twisting).

\subsection{Numerical results in the general case}\label{sec: general_case}

The special cases for the Andreev spectrum discussed in Sec.~\ref{sec:specialcases} and Appendix~\ref{sec: qualitative_explanation_of_spectrum} ($\theta=0$, $\theta=\pi$, $\Phi=0$, and $\Phi_0=\Phi_0$), correspond to the tiles composing the outer frames of panels~(a)--(d) in Fig.~\ref{fig: andreev_levels_current_phase}. In the general case, corresponding to intermediate values of $\theta$ and $\Phi$, we could not further simplify the general analytical expression of Eq.\ (\ref{eq: analytical expression for spectrum}). At the same time, the numerical results of Figs.~\ref{fig: andreev_levels_current_phase}(a) and~\ref{fig: andreev_levels_current_phase}(c) demonstrate that qualitatively, the behavior of the Andreev levels
is similar to that for the special cases.

Generally, the decrease of angle $\theta$ between the filters' magnetizations suppresses transport of split Cooper pairs. As a result, the Andreev levels become flatter, completely loosing dependence on $\varphi$ in the case of parallel filters.

Similarly to the situation discussed in Appendix~\ref{sec: qualitative_explanation_of_spectrum}, the levels can cross at zero energy [see cusps at $E=0$ for the positive part of the spectrum in Fig.~\ref{fig: andreev_levels_current_phase}(c)]. This is possible due to the BdG symmetry (\ref{eq: fundumental symmetry of BdG Hamiltonian}) of $H_\mathrm{BdG}$, which implies that the matrix element of the Hamiltonian between a positive-energy state $\vert\psi\rangle$ and its negative-energy BdG partner $\mathcal{P}\vert\psi\rangle$ is zero, $\langle \psi\vert H_\mathrm{BdG}\vert\mathcal{P} \psi\rangle=0$. Therefore a (positive) level does not repel from its (negative) mirror image, and the levels can cross at $E=0$.

At the same time, level crossing at $E=0$ is not universally protected by the BdG symmetry. For example, at $\theta=\pi$, the levels are spin-degenerate, and an energy level $E$ can repel not from its BdG partner with energy $-E$, but from a physically different state with opposite spin (still corresponding to $-E$). This leads to avoided level crossing at $E=0$.

\section{Josephson current}\label{sec: josephson_current}

\subsection{Current-phase relation}

At zero temperature, the current-phase relation $I(\varphi)$ of the SQUID is given by Eq.\ (\ref{eq: current-phase_through_spectrum}). In the short-junction limit, only two Andreev levels determined by Eq.\ (\ref{eq: analytical expression for spectrum}) [see examples shown in Figs.~\ref{fig: andreev_levels_current_phase}(a) and~\ref{fig: andreev_levels_current_phase}(c)], contribute to the sum in the right-hand side of Eq.\ (\ref{eq: current-phase_through_spectrum}).

Dependence of the current-phase relation $I(\varphi)$ on angle $\theta$ between magnetizations of the spin filters and on magnetic flux $\Phi$ strongly varies with geometrical parameters $\alpha$ and $L$ [compare panels~(b) and~(d) of Fig.~\ref{fig: andreev_levels_current_phase}] due to interferential nature of Cooper pair transport in ballistic systems.

When the filters are parallel ($\theta=0$), the supercurrent is absent, $I(\vp)=0$. For the special cases of $\theta=\pi$ or $\Phi=\Phi_0$, the current-phase relation has the same form as for the QPC Josephson junction \cite{Beenakker 1991}. At $\theta=\pi$, it reads
\begin{equation} \label{ISQPC}
I(\vp) = \frac{e\Delta}{2\hbar} \frac{T_0 \sin\vp}{\sqrt{1-T_0 \sin^2 (\vp/2)}},
\end{equation}
with transparency $T_0$ given by Eq.\ (\ref{T0}), while at $\Phi=\Phi_0$, the form is the same but with $\Delta$ and $T_0$ replaced by $\Delta_\mathrm{eff}$ and $T_\mathrm{eff}$ [Eqs.\ (\ref{Deltaeff}) and (\ref{Teff})], respectively.

For arbitrary values of $\theta$ and $\Phi$, we can distinguish two essentially different types of the $I(\varphi)$ behavior depending on the presence or absence of Andreev level crossing at $E=0$.

Panel~(b) of Fig.~\ref{fig: andreev_levels_current_phase} illustrates the case when the Andreev levels do not cross zero energy. The current-phase relation in this case is qualitatively similar to that of the QPC junction.

Panel~(d) of Fig.~\ref{fig: andreev_levels_current_phase} illustrates the situation when the Andreev levels do cross zero energy. The crossings [represented by cusps at $E=0$ for the positive part of the spectrum in Fig.~\ref{fig: andreev_levels_current_phase}(c)] result in discontinuities of the current-phase curves. With decreasing $\theta$, the concave part of the lower Andreev level is replaced by a convex segment. When this level provides the main contribution (see, e.g., the case of $\theta=0.6\pi$ and $\Phi=0$), this change of curvature implies switching between $0$-junction-type (concave Andreev level) and $\pi$-junction-type \cite{GolubovReview2004,BuzdinReview2005} (convex Andreev level) segments of the $I(\varphi)$ curve. At sufficiently small $\theta$, the crossing points can shift to $\varphi=\pm\pi$ and disappear, then the systems can becomes a pure $\pi$ junction. In Fig.~\ref{fig: andreev_levels_current_phase}(d), this happens, e.g., in the case of $\theta=0.4\pi$ and $\Phi=0$. The upper Andreev level in this particular case supports the tendency to a $\pi$ junction and enhances the contribution of the lower level.
At the same time, a $\pi$ junction can also be achieved in more complicated situations when the upper level demonstrates the opposite (e.g., $\theta=0.2\pi$ and $\Phi=0.4\Phi_0$) or mixed (e.g., $\theta=0.4\pi$ and $\Phi=0.2\Phi_0$) behavior.

\subsection{Critical current}
\begin{figure*}
\begin{subfigure}
\centering{\includegraphics[width=0.3\linewidth]{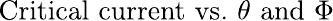}}
\end{subfigure}
\vspace{5mm}

\begin{subfigure}
\centering{\includegraphics[width=0.45\linewidth]{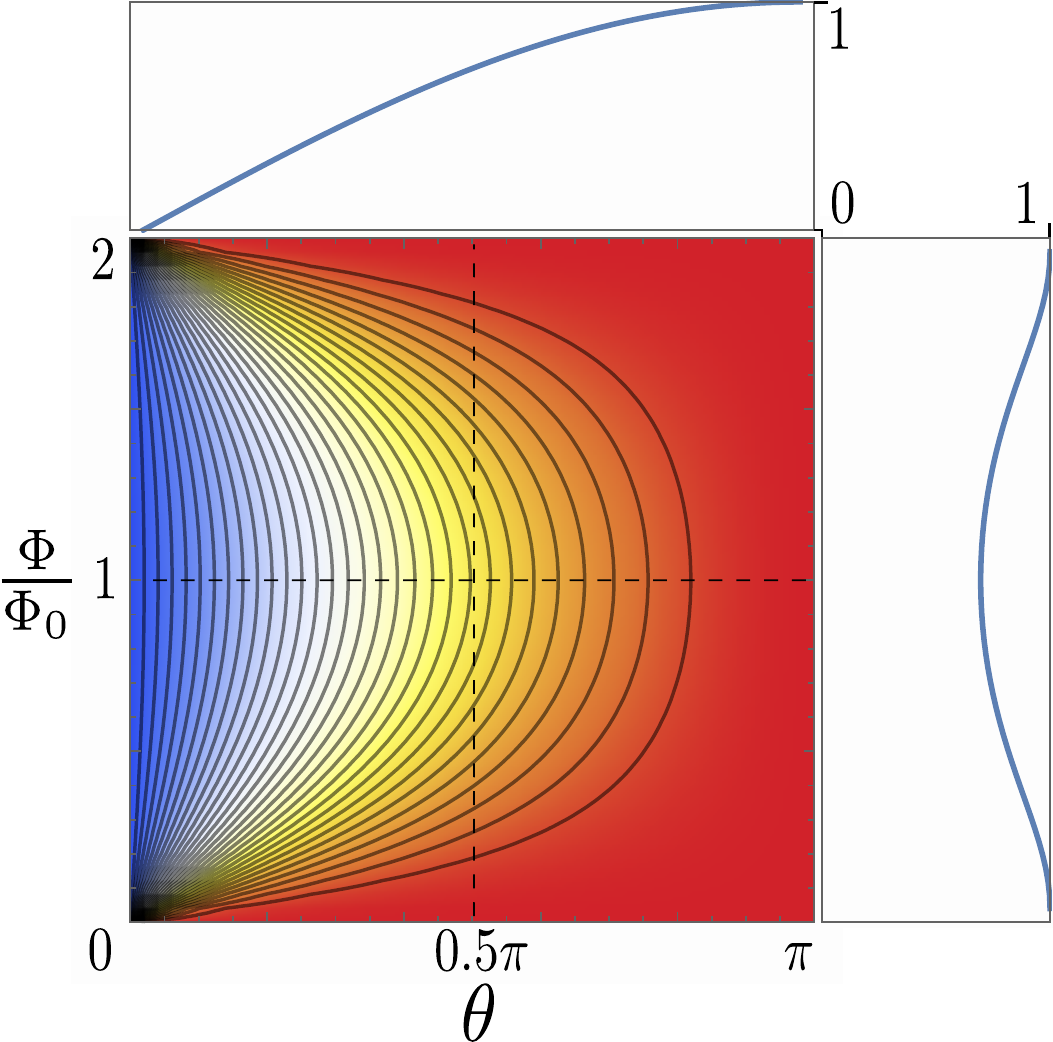}}
\end{subfigure}
\hfill
\begin{subfigure}
\centering{\includegraphics[width=0.45\linewidth]{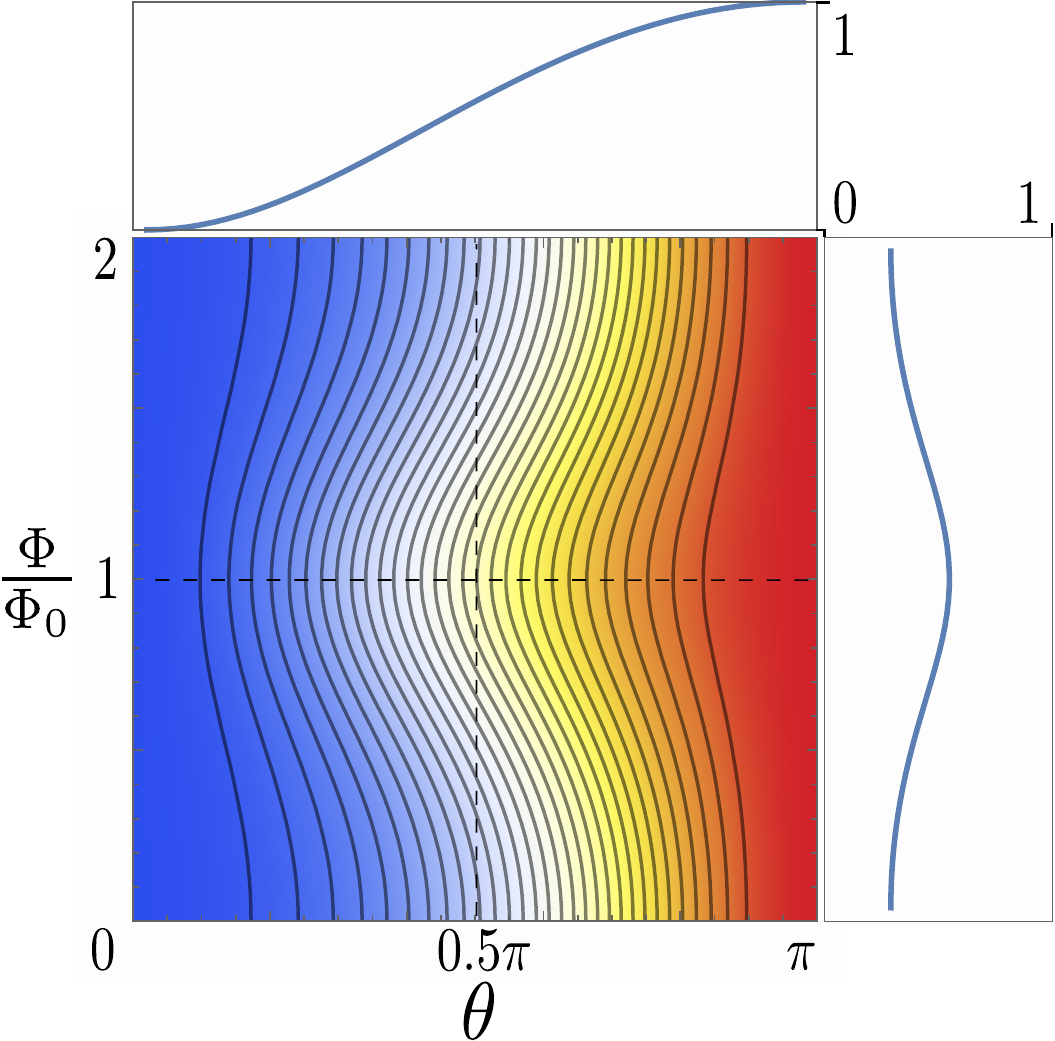}}
\end{subfigure}
\vspace{2mm}

\begin{subfigure}
\centering{\includegraphics[width=0.45\linewidth]{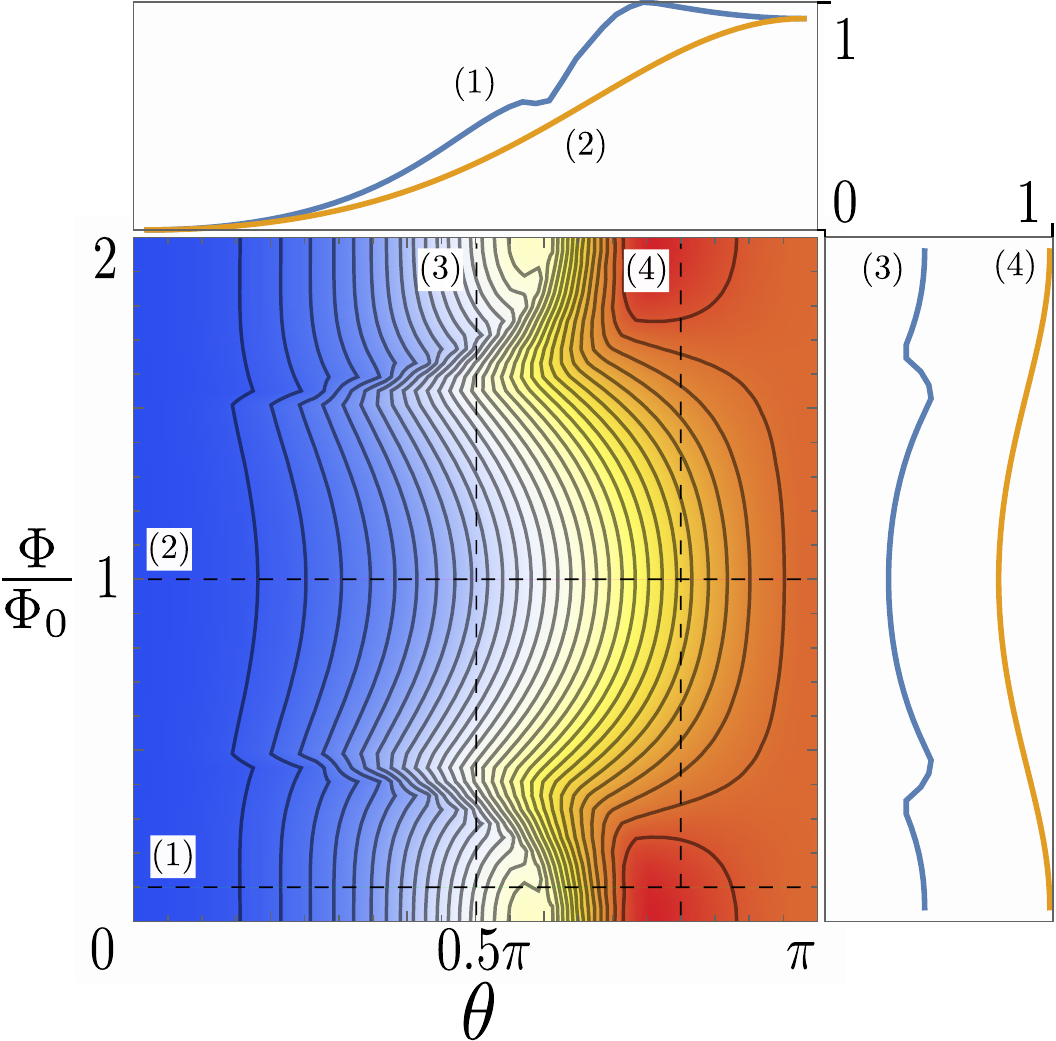}}
\end{subfigure}
\hfill
\begin{subfigure}
\centering{\includegraphics[width=0.45\linewidth]{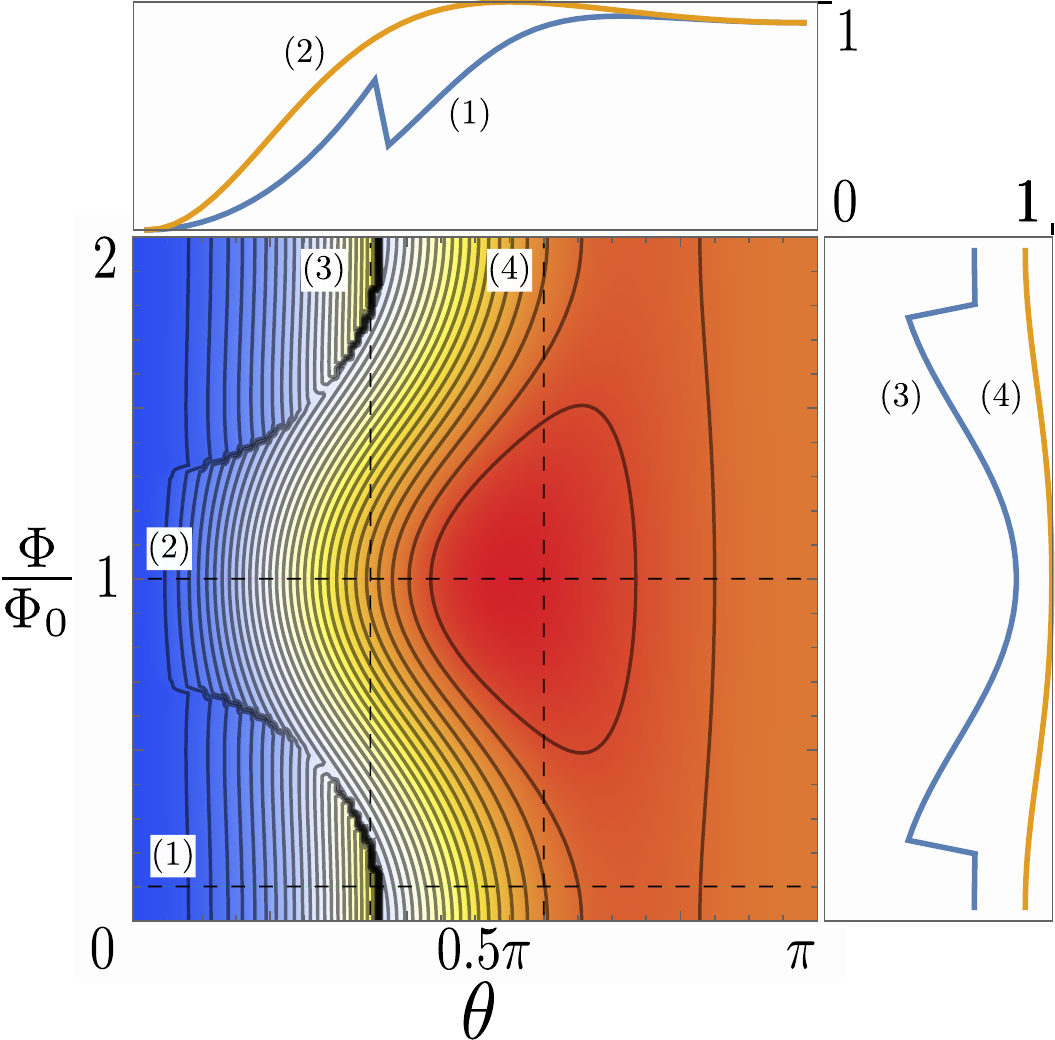}}
\end{subfigure}
\vspace{2mm}

\begin{subfigure}
\centering{\includegraphics[width=0.4\linewidth]{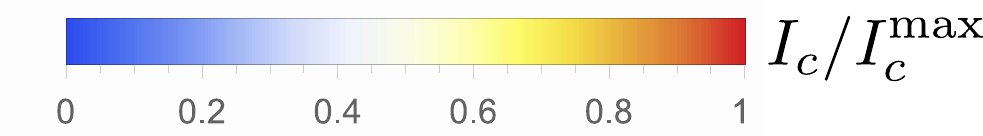}}
\end{subfigure}
\caption{
Dependence of critical current $I_c$ on angle $\theta$ between the magnetization directions of the spin filters (horizontal axis in each panel) and magnetic flux $\Phi$ (vertical axis in each panel) for different geometries ($\alpha$ and $L$ parameters). Color shows the critical current normalized to the maximal value for the corresponding geometry, $I_c/I_{c}^{\mathrm{max}}$. Outsets demonstrate behavior of $I_c$ along the cross sections shown in the main plots. The upper panels (left: $\alpha=\pi$, $L=0$; right: $\alpha=0.6\pi$, $L=0$) illustrate relatively simple behavior: $I_c$ monotonically grows with $\theta$ and monotonically varies with $\Phi$ on the $\left[0,\Phi_0\right]$ interval, reaching either minimum or maximum at $\Phi=\Phi_0$. However, for some geometries, the $I_c$ behavior is much more complicated. The lower panels (left: $\alpha=0.4\pi$, $L=0.5\pi$; right: $\alpha=0.1\pi$, $L=0.3\pi$) illustrate nonmonotonic dependence of $I_c$ on $\theta$ and $\Phi$.}
\label{fig: crit_current_maps}
\end{figure*}

Explicit analytical results for the critical current $I_c$ are available in simple special cases. When the filters are parallel ($\theta=0$), the supercurrent is absent and $I_c=0$. In the cases of $\theta=\pi$ or $\Phi=\Phi_0$, the critical current corresponds to the QPC-type form of the current-phase relation, Eq.\ (\ref{ISQPC}) \cite{Beenakker 1991}. At $\theta=\pi$, we have
\begin{equation}\label{eq: crit_current_QPC}
I_c = \frac{e\Delta}{\hbar} \left( 1- \sqrt{1-T_0} \right),
\end{equation}
with transparency $T_0$ given by Eq.\ (\ref{T0}), while at $\Phi=\Phi_0$, the form of the expression is the same but with $\Delta$ and $T_0$ replaced by $\Delta_\mathrm{eff}$ and $T_\mathrm{eff}$ [Eqs.\ (\ref{Deltaeff}) and (\ref{Teff})], respectively.

At arbitrary $\theta$ and $\Phi$, the critical current can be found numerically.
Due to interferential nature of Cooper pair transport in ballistic systems, behavior of $I_c$ in the $\theta$-$\Phi$ plane strongly depends on the geometrical parameters $\alpha$ and $L$. Several representative examples are shown in Fig.~\ref{fig: crit_current_maps}.

The upper left panel illustrates the simplest case. The critical current monotonically decreases with decreasing $\theta$, reaching zero when the magnetizations are parallel. Dependence on $\Phi$ resembles the usual SQUID behavior, but with nonzero minimal current (at $\Phi=\Phi_0$) and with doubled periodicity. The upper right panel shows ``inverted'' dependence on magnetic flux: $I_c$ is maximal at $\Phi=\Phi_0$.

The lower panels present more complicated regimes. The critical current in these cases depends nonmonotonically on $\theta$. Moreover, the $I_c(\Phi)$ dependence may be nonmonotonic in the $\left[0,\Phi_0\right]$ interval.

We can compare our results for the dependence of the critical current vs.\ misorientation angle, $I_c(\theta)$, with analogous results of Ref.\ \cite{Barash2002}, where the S-FIF-S structure was considered.
In the latter structure, the two ferromagnets are connected in series (while our system has the parallel connection). The distinguishing feature of our results is the possibility of several extrema of $I_c(\theta)$ in the $\left[ 0,\pi\right]$ interval (in contrast to a single minimum in the figures presented in Ref.\ \cite{Barash2002}).

\section{Discussion}\label{sec:discussion}

The main goal of our theoretical consideration was to elucidate the mechanisms of Josephson transport due to split Cooper pairs in ballistic spin-filtering SQUIDs. In order to illuminate distinctive physical features of such processes, we have made a number of simplifying assumptions, the most crucial of them being high-transmission single-channel interferometer arms in the short-junction limit and half-metallic ferromagnets as spin filters. These assumptions are challenging for direct experimental realization of the considered structures. Still, recent experimental progress makes it possible to approach our theoretical limit.

Experimentally, a highly transmissive single-channel interferometer loop can be realized in semiconducting structures \cite{Datta1985}. While being a challenging task, good-quality interfaces between superconductors and a two-dimensional electron gas in semiconducting heterostructures and quantum wells have been experimentally realized \cite{Eroms2005,Batov2007,Kononov2016}. An interesting possibility to create the spin-filtering regions is to employ proximity-induced exchange coupling that can be controlled electrically, similarly to what is being done in the context of spintronics applications \cite{Spiesser_arXiv2017,Zollner_arXiv2017}.

Alternatively, the interferometer loop can be implemented with the help of single-crystalline metallic nanowires obtained by templated electrodeposition \cite{Ryazanov_private}. Single Josephson junctions through such nanowires have already been realized \cite{Skryabina2017}, and the developed technique opens up the possibility to fabricate nanowire-based complex hybrid structures consisting of normal and ferromagnetic parts \cite{StolyarovPatent}.
The next step in this direction could be substituting sections of conventional ferromagnet with sections of a half-metallic ferromagnet in order to achieve absolute spin filtering. Among possible candidates \cite{Pickett2001,Coey2002}, half-metallic CrO$_2$ \cite{Keizer2006,Anwar2010,Singh2015} is the best studied material for superconducting heterostructures. Moreover, CrO$_2$ can be realized in the form of single-crystalline nanowires \cite{Zou2008,Zhao2011}, and good interfaces to such nanowires can be achieved in Josephson structures \cite{Singh_arXiv2016}.

Although we model beam splitting with the help of three-terminal Y-form junctions, this specific geometry is not critical for the physics that we discuss. Our Y splitters model possibility of crossed Andreev reflection. Experimentally, this effect could also be achieved in the case of straight nanowires \cite{Skryabina2017,StolyarovPatent},
if they are attached to a superconductor with separation smaller than the coherence length \cite{BWL2004,BL2005}. Regarding the quality of interfaces, we note that single-crystalline Au nanowires can form nearly perfect interfaces with superconducting Al \cite{Jung2011}.

Taking Al
for the superconducting reservoirs and single-crystalline Cu \cite{Skryabina2017,StolyarovPatent} or Au \cite{Jung2011} for the normal-metallic wires, one can achieve the coherence length in the wires of the order of several microns, so that the short-junction limit can really be reached in SQUID structures.
At the same time, as follows from Eq.\ (\ref{eq: crit_current_QPC}), the critical current in the SQUID can be as high as the critical current in a single-channel QPC Josephson junction. In the case of Al superconducting reservoirs, the latter can reach the order of several tens of nanoamperes, which can be reliably measured with the help of currently available experimental techniques.

\section{Conclusions}\label{sec: conclusion}

We have calculated the Andreev levels and the corresponding Josephson current in ballistic SQUID with spin filtering inside half-metallic ferromagnetic arms (Fig.~\ref{fig: model_sketch}) as a function of two control parameters, angle $\theta$ between the magnetizations of the spin filters and external magnetic flux $\Phi$. The transport of Cooper pairs in the presence of absolute spin filters is entirely due to split-pair processes, with two electrons passing through different interferometer arms. The arms were assumed to be highly transmissive single-channel wires in the short-junction limit.

Technically, we employ the scattering matrix approach, starting with derivation of the general analytical result (\ref{eq: analytical expression for spectrum}) for the Andreev levels at arbitrary energy-independent nonsuperconducting scatterer between the superconducting reservoirs. Due to the short-junction limit, the supercurrent is carried by exactly two Andreev bound states (which can be degenerate in limiting cases).

The obtained expression for the Andreev levels is then applied in the special cases of $\theta=0$, $\theta=\pi$, $\Phi=0$, and $\Phi=\Phi_0$, where analytical progress is possible. In particular, in the $\theta=\pi$ case, the spectrum turns out to be insensitive to $\Phi$. Putting $\Phi=0$ for simplicity, we then find that spin symmetry is restored, spin is conserved, and the SQUID becomes equivalent to the QPC Josephson junction with transparency determined by geometrical parameters of the system (phases acquired by quasiparticles in the beam splitter and in the spin-filtering insets). In the $\Phi=\Phi_0$ case, ``flipped spin'' is conserved (the flipped spin sectors are composed of opposite spins near opposite superconductors), and the system again reduces to the QPC Josephson junction with effective transparency. However, in this case, the order parameter in the QPC formulas is effectively reduced in comparison to $\Delta$ of the reservoirs.

Different geometrical parameters of the system lead to qualitatively different behavior of the SQUID characteristics (the Andreev levels, the current-phase relation, and the critical Josephson current) as a function of $\theta$ and $\Phi$. The current-phase relation can change its amplitude and shape, in particular, varying between $0$- and $\pi$-junction form. The transition goes through intermediate states, in which $I(\varphi)$ is composed of $0$- and $\pi$-type segments with jumps between them.

As a result, the critical current $I_c$ can become a nonmonotonic function of $\theta$ (as the angle varies between the parallel and antiparallel configuration). Periodicity with respect to the magnetic flux is $2\Phi_0$, i.e., doubled, in comparison to conventional SQUIDs. A simple process, in which two electrons of a Cooper pair pass through two different arms, is not sensitive to $\Phi$ at all. However, the scattering matrix approach effectively sums up all possible trajectories, and the doubled periodicity is actually due to more complicated processes, in which one electron simply passes through its arm, while the second one passes through the other arm and also makes an additional loop inside the nonsuperconducting part of the interferometer. Due to interference effects, $I_c$ can acquire inverted dependence on $\Phi$ (with maximum shifted by half-period, i.e., from $0$ to $\Phi_0$) or even become a nonmonotonic function between $0$ and $\Phi_0$ (half-period).

\acknowledgments
The idea of this research was formulated in the course of our conversations with V.~V.\ Ryazanov, and we are grateful to him for useful discussions of our results. We also acknowledge useful discussions with I.~S.\ Burmistrov, E.~V.\ Deviatov, M.~V.\ Feigel'man, D.~A.\ Ivanov, S.~V.\ Mironov, and V.~S.\ Stolyarov. This work was supported by the Russian Science Foundation (Grant No.\ 16-42-01035).

\appendix

\section{Interferential opacity of splitters}\label{App: interferential opacity of splitters}

At $\alpha=0$, destructive interference generally leads to complete opacity of the system (except for special sets of parameters). Here, we explain this effect in the simple case of antiparallel orientation of the spin filters, $\theta=\pi$.

In order to show that the effect is not limited to ideal (nominally reflectionless) splitters with $Y_{33}=0$ (defined in Sec.~\ref{sec: S-matrices of beam splittres}), we consider a more general matrix:
\begin{equation}\label{eq: generalized_T_splitter_matrix}
Y=\begin{pmatrix}
-\sin^2\frac{\gamma}{2} & \cos^2\frac{\gamma}{2} &\frac{1}{\sqrt{2}}\sin\gamma \\
\cos^2\frac{\gamma}{2} & -\sin^2\frac{\gamma}{2} &\frac{1}{\sqrt{2}}\sin\gamma \\
\frac{1}{\sqrt{2}}\sin\gamma &\frac{1}{\sqrt{2}}\sin\gamma & -\cos\gamma
\end{pmatrix},
\end{equation}
where the $\gamma$ parameter characterizes reflection from the splitter for quasiparticles coming from terminal 3 (from reflectionless splitter at $\gamma=\pi/2$ to completely detached terminal 3 at $\gamma=0$).

\begin{figure}[t]
\includegraphics[width=\columnwidth]{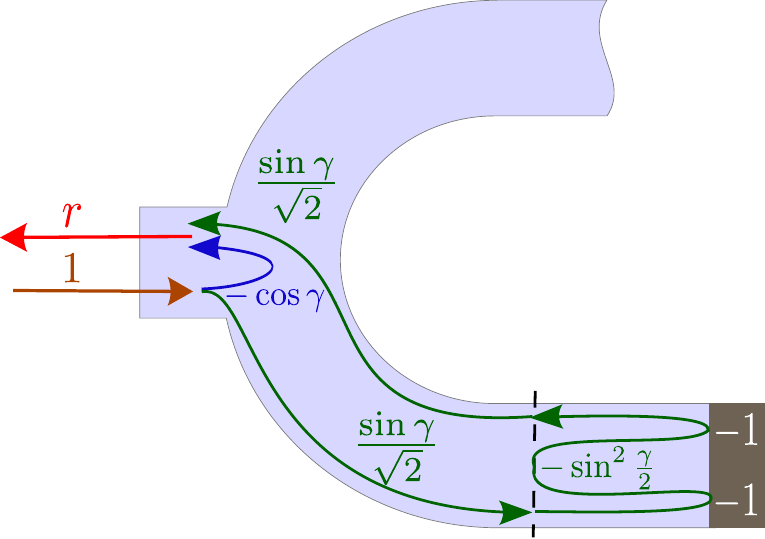}
\caption{
Explanation of effective interferential opacity of the beam splitter in the simple case of $\theta=\pi$. The brown arrow shows the incoming quasiparticle and the red one is the reflected part. The blue line denotes immediate reflection by non-ideal beam splitter. The green arrow corresponds to the path inside the dead-end ``resonator'' formed by the spin filter (oriented oppositely to the quasiparticle spin) and the beam splitter.
}
\label{fig: interferential opacity}
\end{figure}

At $\theta=\pi$, for quasiparticles with spin up along the $x$ axis, the upper arm of the interferometer is transparent while the lower arm is effectively interrupted in the middle (due to the impenetrable spin filter), see Fig.~\ref{fig: interferential opacity}. At first glance, the quasiparticle can pass the interferometer through the upper arm. However, the interference processes involving the ``dead end'' (the lower arm) actually block this passage \cite{Gefen1984}. In order to demonstrate this, we calculate the reflection amplitude $r$ for a quasiparticle trying to enter the splitter from the left (terminal 3).

The reflection amplitude consists of two contributions.
First, the quasiparticle can be reflected immediately when trying to enter the splitter (blue line in Fig.~\ref{fig: interferential opacity}). The corresponding amplitude follows directly from Eq.\ (\ref{eq: generalized_T_splitter_matrix}):
\begin{equation}
r_1=-\cos\gamma.
\end{equation}
Second, the quasiparticle can enter the dead end (green line in Fig.~\ref{fig: interferential opacity}). Then it experiences multiple reflections from the impenetrable wall and the splitter and finally can be reflected back to the left. The corresponding contribution is the following sum:
\begin{multline}
r_2=\frac{\sin\gamma}{\sqrt{2}}\left[-1+(-1)\left(-\sin^2\frac{\gamma}{2}\right)(-1)+\dots\right]\frac{\sin\gamma}{\sqrt{2}}
\\
=\cos\gamma-1.
\end{multline}
The total amplitude is then
\begin{equation}
r=r_1+r_2=-1.
\end{equation}

In this way, for spin-up (along the $x$ axis) quasiparticles even the first beam splitter of the interferometer effectively acts as an infinitely high barrier [note that our argument is independent on the second (right) beam splitter]. The same reasoning is applicable for spin-down quasiparticles. Consequently, the whole interferometer loop is impenetrable although there is an absolutely transparent arm for each spin.

This simple example shows that interference is one of the crucial factors determining the behavior of the system.

\section{Qualitative explanation of spectrum shape for $\Phi=0$}\label{sec: qualitative_explanation_of_spectrum}

\begin{figure*}
\includegraphics[width=\textwidth]{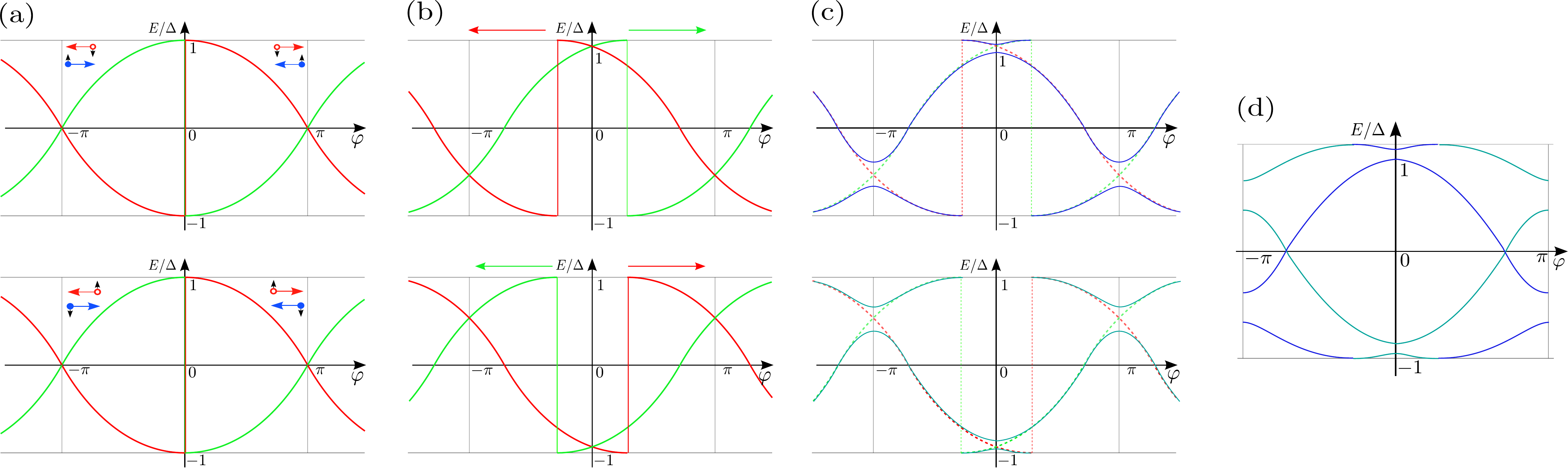}
\caption{Qualitative explanation of the Andreev spectrum shape for $\Phi=0$.
The upper and lower rows correspond to two different spin sectors.
The red and green curves in each case are the Andreev levels of opposite chirality [the blue and red arrows in column~(a) show the direction of motion for electrons and holes, respectively].
Column~(a): transparent spin-independent scatterer.
Column~(b): shifting of the levels due to different phases in different spin sectors.
Column~(c): avoided level crossings due finite transparency, which causes mixing between the states of different chirality.
Column~(d): the overall structure of the Andreev levels.}
\label{fig: qualitative_explanation_spectrum}
\end{figure*}

In this appendix, we discuss how characteristic behavior of the Andreev spectrum at zero magnetic flux (see Fig.~\ref{fig: andreev_levels_current_phase}, $\Phi=0$) can be understood qualitatively.

First, we note that since the $z$ component of spin is conserved at $\Phi=0$ (see Sec.~\ref{sec: zero_flux}), the spectrum consists of two independent spin sectors: `up-spin electron plus down-spin hole', and the sector with opposite spins.

Second, if we start from the case of $\theta=\pi$, the scattering by the interferometer becomes trivial with respect to spin (see Sec.~\ref{sec: antiparallel_magnetizations}), therefore the Andreev levels for the two spin sectors coincide. The spectrum is then described by effective transparency, see Eqs.\ (\ref{eq: spectrum of short junction}) and (\ref{T0}).

Third, if we additionally assume, for instance, $\alpha=\pi$ and $L=0$, the junction becomes perfectly transparent.
In the transparent case, the eigenstates are characterized by the quantity that can be called ``chirality'': for an electron moving to the right, the Andreev-reflected hole moves to the left. The state, in which an electron moves to the left, has the opposite chirality.

The spectrum structure in this simplest case (equivalent to the perfectly transparent QPC Josephson junction) is shown in Fig.~\ref{fig: qualitative_explanation_spectrum}(a).
For arbitrary $\theta$, $\alpha$, and $L$, the spectrum is changed by two main factors:

(i) If $S_\mathrm{interf}$ becomes nontrivial in the spin space [while still conserving the $z$ spin projection, which is true for $\Phi=0$, see Eq.\ (\ref{eq: S-matrix for zero external flux})], electrons with different spins accumulate different phases when traversing the interferometer in the same direction. In Eq.\ (\ref{eq: S-matrix for zero external flux}), this difference is represented by $\Psi$. As a result, the energy levels shift as functions of $\varphi$ (similarly to the case of Ref.~\cite{Kuplevakhskii1990}). Since the two levels of the same spin sector have opposite chirality,
they shift in opposite directions: apart from or towards each other, depending on the spin sector, as shown in Fig.~\ref{fig: qualitative_explanation_spectrum}(b).

(ii) Normal reflection from the interferometer (in contrast to the Andreev reflection from the SN boundaries) leads to mixing of states with different chirality (belonging, at the same time, to the same spin sector). As a result, the corresponding avoided level crossings appear (note that states from different spin sectors are not mixed and the levels still cross). This is illustrated in Fig.~\ref{fig: qualitative_explanation_spectrum}(c). In Eq.\ (\ref{eq: S-matrix for zero external flux}), reflection is represented by $\rho_{\uparrow}$ and $\rho_{\downarrow}$.

The full spectrum is then obtained as combination of the two spin sectors, as shown in Fig.~\ref{fig: qualitative_explanation_spectrum}(d).

Analyzing the scattering matrix in the case of $\Phi=0$, Eq.\ (\ref{eq: S-matrix for zero external flux}), we see that the only feature not taken into account in the simplified description leading to Fig.~\ref{fig: qualitative_explanation_spectrum}, is the difference of the probabilities for reflection (and transmission) of two different spins. Nevertheless, our description is able to reproduce (at least, qualitatively) the main types of the $E(\varphi)$ dependence, see Figs.~\ref{fig: andreev_levels_current_phase}(a) and~\ref{fig: andreev_levels_current_phase}(c).

\end{document}